\documentclass[preprint,aps, preprintnumbers, nofootinbib,showpacs]{revtex4}
\usepackage[dvips]{graphics}

\newcommand{\gsim}[2]{
\setlength{\unitlength}{12pt}
\begin{picture}(1.4,1.)
\put(.7,-0.3){\makebox(0.0,1.)[t]{$>$}}
\put(.7,-0.3){\makebox(0.0,1.)[b]{$\sim$}}
\end{picture}#2}
\begin{document}
\preprint{IRB-TH-20/05}

\title{Generalized holographic dark energy and the IR cutoff problem}

\author{B. Guberina\footnote{guberina@thphys.irb.hr}}
\affiliation{\footnotesize Rudjer Bo\v{s}kovi\'{c} Institute,
         P.O.B. 180, 10002 Zagreb, Croatia}

\author{R. Horvat\footnote{horvat@lei3.irb.hr}}
\affiliation{\footnotesize Rudjer Bo\v{s}kovi\'{c} Institute,
         P.O.B. 180, 10002 Zagreb, Croatia}

\author{H. Nikoli\' c\footnote{hrvoje@thphys.irb.hr}}
\affiliation{\footnotesize Rudjer Bo\v{s}kovi\'{c} Institute,
         P.O.B. 180, 10002 Zagreb, Croatia}

\begin{abstract}

We consider a  holographic dark energy model, in which both the
cosmological-constant (CC) energy density $\rho_{\Lambda}$ 
and the Newton constant $G_N$ are varying quantities, to study the
problem of setting an effective field-theory 
IR cutoff. Assuming that ordinary matter scales
canonically, we show that the continuity equation univocally fixes the IR 
cutoff,
provided a law of variation for either $\rho_{\Lambda}$ or $G_N$ is known. 
Previous
considerations on holographic dark energy 
disfavor the Hubble parameter as a candidate for the IR
cutoff (for spatially flat universes), since in this case the ratio of
dark energy to dark matter is not allowed to vary, thus hindering a
deceleration era of the universe for the redshifts $z \gsim \; 0.5$. 
On the other hand, the
future event horizon as a choice for the IR cutoff is being favored in 
the literature, although 
the `coincidence problem' usually cannot be 
addressed
in  that case. We extend considerations to spatially curved universes,
and show that with the Hubble parameter as a choice for the IR cutoff one
always   
obtains a universe that never accelerates or a universe that
accelerates all the time, thus making  
the transition from deceleration to acceleration impossible.
Next, we apply the IR  cutoff consistency procedure to a 
renormalization-group (RG) running CC model, 
in which the low-energy
variation of the CC is due to quantum effects of particle fields having 
masses near the Planck scale. We show that bringing such a model (having the
most general cosmology for running CC universes) 
in full
accordance with holography 
amounts to having such an IR cutoff which scales as a
square root of the Hubble parameter. We find that such a setup, in which the
only undetermined input represents the true ground state of the vacuum, 
can give 
early deceleration as well as late time acceleration. The possibility of
further improvement of the model is also briefly indicated.        

\end{abstract}
\newpage

\maketitle

A few years ago, Cohen {\it at al.} \cite{1}, motivated by the 
Bekenstein bound  \cite{2} on the maximal possible entropy, showed that   
systems which did not contain black holes might still allow a 
self-description by
means of conventional quantum field theory, if a certain relationship between
the UV and IR cutoffs was obeyed. Applying this limit \cite{1} when the
volume under consideration  is the universe itself, results in one of the
most elegant solutions to the (`old') cosmological constant (CC) problem
\cite{3}. The limit on the zero-point energy density $\rho_{\Lambda}$ 
in \cite{1} represents a more stringent version of the
holographic principle \cite{4, 5}. In short, such a principle states that in 
the presence of quantum
gravity, 
all of the information
contained in a certain volume of space 
can be represented by a theory that counts
degrees of freedom only on the boundary of that region.

Recently, starting from \cite{6}, a strong impetus for a more thorough
investigation of the  Cohen {\it at al.}  bound \cite{1} and its consequences  
in cosmology was
triggered by a widespread consensus about the presently accelerated
expansion of the universe \cite{7}. In addition, since the size of the
region (providing an IR cutoff) is varying in an expanding universe,
$\rho_{\Lambda}$ (usually dubbed the holographic energy density) 
is promoted to a dynamical quantity. Therefore one hopes 
this might also shed some light on a puzzling `coincidence problem'
\cite{8}.
One can show that irrespective of the hierarchy between the UV cutoff and
particle masses,
$\rho_{\Lambda}$ generated by vacuum fluctuations is always dominated
by UV modes, and that the limit from \cite{1} can be rewritten as
\begin{equation}
\rho_{\Lambda }(\mu ) \; = \kappa \; \mu^{2} \; G_{N}^{-1}(\mu ) \;,
\end{equation}
where $\mu $ denotes the IR cutoff, and the bound from \cite{1} is
saturated for $\kappa \simeq 1$. The case when  $G_N$ from (1) is also
promoted to a dynamical quantity [as explicated in (1)]  
was considered for the first time in \cite{9}, and
afterwards in \cite{10}. Such a setup we call a generalized holographic dark
energy. By the simplest choice for $\mu $ in the form of the present
Hubble parameter, one indeed obtains the dark energy density very close to
the observed one. Thus, for this choice  
of the IR cutoff, the parameter
$\kappa $ in (1) should be close to unity. 
One thus sees how efficiently holography stabilizes
the vacuum energy and ameliorates the `old' CC problem by eliminating the
need for fine-tuning.  

Most authors modeled dark energy from (1) to evolve independently of matter
density $\rho_m $, i.e. to behave as a perfect fluid. In this case, both 
components
evolve according to the $w$-parameter from their equations of state (EOS). 
Another
modeling of $\rho_{\Lambda}$ from (1) is through interacting fields
\cite{11}, where an interaction between dark energy and matter fields is
postulated; in this setup the components no longer evolve according to
their $w$-parameters. Scaling of $\rho_{\Lambda}$ due to a different kind of
interaction involving  nonstatic $G_N$  
was also investigated \cite{9, 10, 12}. It is easy to see why the
identification of the IR cutoff with the Hubble distance for spatially flat
universes  leads to unsatisfactory cosmologies for all the cases above 
(although the interpretation for each case is different). Indeed, by plugging (1)
in the Friedmann equation for flat space one obtains
\begin{equation}
H^2 = \frac{8 \pi \kappa}{3} \mu^2 (1 + r)\;,
\end{equation}
where $r=\rho_m /\rho_{\Lambda}$. Thus, the choice $\mu  \sim H$ would
require the ratio $r$ to be a constant. This holds irrespective 
of whether a fluid is
perfect or not and irrespective of whether $G_N$ is varying or not. 
The interpretation of
various cases is, however, different. For perfect fluids, $r=const.$ means
that the equation of state for dark energy unavoidably  matches that of
pressureless matter, $w =0$. Thus, we cannot explain the accelerating
expansion of the present universe. For interacting fluids, one usually can 
generate accelerated expansion with $r=const.$ as well as ameliorate the
`coincidence problem', but fails to explain that the acceleration era set
in just recently and was preceded by a deceleration era at $z \gsim ; 0.5$
(see, however, \cite{11}).

Another studied choice for the IR cutoff was the particle horizon distance
\cite{13}. However, this option fails for perfect fluids because it is
impossible to obtain a $w$-parameter characterizing  accelerated universes.
The most popular choice is by far the future event horizon. This choice
seems to work well both for perfect \cite{13} and interacting fluids
\cite{14}. Within this choice, however, most approaches suffer from the
`coincidence problem'. For related works, see \cite{15}.      
  
In the present paper we adopt the framework consistent with the generalized
holographic energy density. This simplest possibility is to let the matter
energy density  scale canonically $\sim a^{-3}$, i.e. behaving as a perfect
fluid. The scaling of $\rho_{\Lambda}$ is then induced by the scaling of $G_N$
(and {\it vice versa} ) and is described by the equation of continuity of the
form
\begin{equation}
\dot{G}_{N}(\rho_{\Lambda } + \rho_m ) + G_N \dot{\rho }_{\Lambda } = 0  \;,
\end{equation}
where overdots denote time derivatives. Eq. (3) means that 
$G_{N} T_{total}^{\mu \nu}$ 
and $T_{matter}^{\mu \nu }$ are separately conserved \cite{fot1}.
The difference
of the framework phrased by (3) with respect to other models employing holographic 
energy density is twofold. Firstly, the scaling of $\rho_{\Lambda }$ is not
caused by the energy transfer with the matter component, but goes at the 
expense of the time-dependent
gravitational coupling \cite{fot2}.
Secondly, $\rho_{\Lambda }$ in (1)-(3) represents the
variable (or interacting) but `true' CC, with the EOS
$w_{\Lambda} \equiv p_{\Lambda}/{\rho_{\Lambda}}$ being precisely -1.
Although holographic dark energy is usually modeled in the literature
through scalar fields (being in the form of both perfect and interacting 
fluids), we note
that the bound for $\rho_{\Lambda}$ from \cite{1} was derived by considering vacuum 
fluctuations (zero point energies). Since these represent the `true' CC, 
there is no need for invoking a light scalar field with its potential. Note
also that even the
static $\rho_{\Lambda}$ can fit within the bound (1) if $G_N$ is varying.          

The aim of the present paper is threefold. First, we obtain a relation which
univocally fixes the IR cutoff, provided our framework  is phrased by Eqs.
(3) and (1). Thus, if, e.g. the law for $G_N(\mu )$ is specified
[or for $\rho_{\Lambda}$, which is the same as both quantities are connected
through (1)], then the IR
cutoff $\mu $ is no longer liable to a free choice, but is fixed. This is 
in clear contradistinction with the case when the energy transfer is between
$\rho_{\Lambda}$ and $\rho_m $. In this case, the equation of continuity is
of the form
\begin{equation}
\dot{\rho }_{\Lambda } + \dot{\rho }_m + 3H\rho_m = 0  \;.
\end{equation}
We see from (4) that any deviation from the canonical scaling $\rho_m \sim
a^{-3}$ depends decisively on the choice of $\mu $. Hence, with different
choices for $\mu $ one modulates the interaction between $\rho _{\Lambda }$
and $\rho_m $ as described  by (4). Our second step is to set the IR 
cutoff in the scale-fixing relation at
$\mu = H$ without {\it a priori} specifying a law for $\rho_{\Lambda}$, 
in order to
investigate if this choice still remains a bad one for curved universes.
Finally, we study  compatibility of generic variable CC models with the
holographic prediction (1). It is known that such models lead to successful
cosmologies when the transfer of energy is of the type as given by (4).
However, merging such models with the prediction of gravitational
holography leads them to be inconsistent with cosmological observation. 
On the example of the most general running CC model, we show
that the situation can substantially be improved if instead the framework
described by (3) is used. We then study observational consequences of such a
model.    
 
The scale-fixing relation emerges after inserting the holographic prediction
(1) in the equation of continuity (3):
\begin{equation}
\mu = - \frac{G_{N}'(\mu ) \rho_m }{2 \kappa } \;,
\end{equation}
where the prime denotes differentiation with respect to $\mu$. 
(Note that (5) cannot be applied to the case $\rho_m=0$.
In this case, (1) and (3) imply $\dot{\mu}=0$, so
a differentiation with respect to $t$ cannot be expressed
in terms of a differentiation with respect to $\mu$.
Consequently, (5) is not valid in this case.)
If the IR cutoff
represents some energy scale which decreases throughout the 
expansion of the universe,
then from (5) one sees that $G_{N}(t)$ increases as a function of cosmic
time. This raises a possibility of having the gravitational coupling $G_N$ 
which is
asymptotically free, as revealed in some quantum gravity models \cite{17}. 

Setting $\mu = H$ in (5) and $\rho_m = \rho_{m0} a^{-3}$, and after 
combining (1) and (5) with the Friedmann equation for nonzero curvature, 
one obtains the 
following equation for the time evolution of the scale factor $a(t)$:
\begin{equation}\label{eq..}
\dot{a}^2-\gamma a\ddot{a}=\eta,
\end{equation}
where $\gamma$ and $\eta$ are constants. The constant
$\eta$ is proportional to $\Omega_{k0}\equiv -k/H_0^2$, while $\gamma$
does not depend on $\Omega_{k0}$. The signs of these constants depend
on the exact value of $\kappa$.
Eq.~(\ref{eq..}) is a nonlinear second-order differential equation.
In general,
the explicit analytic solutions are difficult to find.
However, below we present an analytical proof that
there are no physically acceptable solutions with the property
that the acceleration $\ddot{a}$ changes the sign during the
evolution. 

First, consider the case $\eta=0$. In this case, the solution
can be found analytically.
Requiring $a(0)=0$,
the solution is
\begin{equation}
a(t)=\left( \frac{t}{t_0} \right)^{\alpha} ,
\end{equation}
where $\alpha\equiv\gamma/(\gamma-1)>0$.
(For $\alpha\leq 0$, the
solution does not correspond to an expanding universe.)
For $0<\alpha<1$, $\ddot{a}<0$ everywhere except at $t\rightarrow\infty$.
For $\alpha=1$, $\ddot{a}=0$ everywhere.
For $1<\alpha<2$, $\ddot{a}>0$ everywhere except at $t\rightarrow\infty$.
For $\alpha=2$, $\ddot{a}=2/t_0^2$ is a constant.
For $\alpha>2$, $\ddot{a}>0$ everywhere except at $t=0$.

Now consider the case $\eta\neq 0$.
If $\ddot{a}$ changes the sign, then there exists a particular
time $t_1$ at which $\ddot{a}(t_1)=0$. Eq.~(\ref{eq..}) then
implies $\dot{a}^2(t_1)=\eta$. This is clearly impossible
for a negative $\eta$, so the rest of the discussion is restricted
to a positive $\eta$. The expanding universe satisfying
$\dot{a}^2(t_1)=\eta$
satisfies the initial condition $\dot{a}(t_1)=\sqrt{\eta}$, so one can
verify that the solution of (\ref{eq..}) reads
\begin{equation}\label{solut}
a(t)=a_1+\sqrt{\eta}(t-t_1),
\end{equation}
where $a_1$ is a free integration constant corresponding to an
arbitrary choice of the initial condition $a(t_1)=a_1$.
Clearly, the solution (\ref{solut}) has the property
$\ddot{a}(t)=0$ for each $t$. Thus, if $\ddot{a}(t_1)=0$
for {\em some} $t_1$, then $\ddot{a}(t)=0$ for {\em all} $t$.
Consequently, $\ddot{a}$ cannot change sign
during the evolution, Q.E.D. \cite{fot3}.

Now we turn to discuss variable CC models in the light of the vacuum decay
law as predicted by gravitational holography and given by  (1). For a 
comprehensive 
list of proposals of various  vacuum
decay laws, see \cite{18}. Cosmological implications of a few mostly
discussed laws in the literature were critically examined in \cite{19}. It
was stressed in \cite{19} that the renormalization-group (RG) running model
of Shapiro {\it et al.} \cite{20, 21} led to the most 
general cosmology for vacuum
decaying universes. The running of $\rho _{\Lambda }$ in \cite{20} is of the
form
\begin{equation}
\rho _{\Lambda }(\mu ) = c_2 \mu^2 + c_0\;,
\end{equation}
where in \cite{20} the RG scale was set at $\mu = H$,  
$c_2 \sim M_{Pl}^2 $ and $c_0 $ was the IR limit of the CC and
represents here the true ground state of the vacuum \cite{fot4}.
The model phrased by
(9) was  based on the observation \cite{22} that even a `true' CC in
conventional field theories could not be fixed to any definite constant
(including zero) owing to the RG running effects. The variation of the CC
arises solely from field fluctuations of the heaviest particles, without
introducing any quintessence-like scalar fields.  Particle contributions to
the
RG  running of the CC which are due to vacuum fluctuations of
massive fields have been properly derived in \cite{23}, with a somewhat
peculiar outcome that more massive fields do play a dominant role in the
running at any scale. The `coincidence problem' is simply understood by
noting that the scale associated to the CC is simply given as a geometrical
mean of the largest and the smallest scale in the universe today. It was
shown \cite{21} that the law (9), in conjunction with the continuity equation 
(4), can
explain recent cosmological observations. In the context of the present
paper, however, we note a serious drawback of the model described by (9) and 
(4),
when trying to accommodate the prediction from holography as given by (1).
Namely, employing (4) implies that $G_N$ is static, and insertion of the law 
(9) in (1) unavoidably sets the ground state of the vacuum to zero $(c_0=0)$.
This means that $\rho_{\Lambda}$ from (9) scales as $H^2 $, leading to
a constant $r$ for spatially flat universes, as explained in detail in the 
first part of the paper. Thus, an attempt to  bring the above model in 
accordance with holography, leads unavoidably  
to undesired phenomenological implications.

Now we try a different approach, in which the model given by (9) is
investigated together with the continuity equation (3). In this case, the
scale $\mu $ is not left to a free choice but is fixed by 
Eq. (5) \cite{fot5}.
From (1), (5) and (9) we find an
explicit expression for the scale $\mu $ in the form
\begin{equation}\label{mu2}
\mu^2=\frac{\sqrt{-c_0 \rho_m} -c_0}{c_2} \;.
\end{equation}
The constants $c_2$ and
$c_0$ can be expressed in terms of the present-day values of the cosmological
parameters as 
\begin{equation}
c_2=\frac{\kappa}{G_0} \frac{1+r_0}{r_0} \;,
\end{equation}
\begin{equation}
c_0=-\frac{\rho_{\Lambda 0}}{r_0} \;.
\end{equation}
Using $\rho_m = \rho_{m0} a^{-3}$, we find from the above expressions that
the ratio $r$ now scales as (for arbitrary spatial
curvature) 
\begin{equation}\label{r}
r=r_0 a^{-3/2}.
\end{equation}
The ratio $r$ becomes larger in the past, so we can expect a transition from
acceleration to deceleration at late times.

By exploiting the Friedmann equation for flat space, one is allowed to
express the scale $\mu $ in the alternative form as 
\begin{eqnarray}
\mu & = & \left(\frac{3}{8\pi\kappa}\right)^{1/4}
\left(\frac{-c_0}{c_2}\right)^{1/4} H^{1/2} \\ \nonumber
& = & 
\sqrt{\frac{3}{8\pi\kappa (r_0+1)}} \sqrt{H_0 H}.
\end{eqnarray}
Hence, if the model described by (9) and (3) respects the holographic
prediction, the scale $\mu $ is fixed at $\mu  \sim H^{1/2}$. 

Solving the Friedmann equation for the scale factor $a(t)$, we obtain the
solution as
\begin{equation}\label{a(t)}
a(t)=r_0^{2/3}\left( e^{\frac{3\beta}{2}t} -1 \right)^{2/3} \;,
\end{equation}
where $\beta \equiv H_0/(1 + r_0 )$. Note that $c_0 < 0$ [Eq.(12)] does not
imply the anti-de Sitter universe, since $G_N $ is also varying
in our model. On
the contrary, (15) shows that the universe in our model is asymptotically de
Sitter, with $a(t) \simeq r_{0}^{2/3} e^{\beta t}$.

Now we are in a position to compare our scale (14), obtained in a consistent
treatment, with the Hubble parameter and the inverse future horizon computed in
the same model. For the most part of the evolution history of the universe
the scales are plotted in Fig. 1. We find it indicative to have our scale
(14) much closer to the inverse future horizon than to the Hubble parameter.
  
\begin{figure}[t]
\centerline{\includegraphics{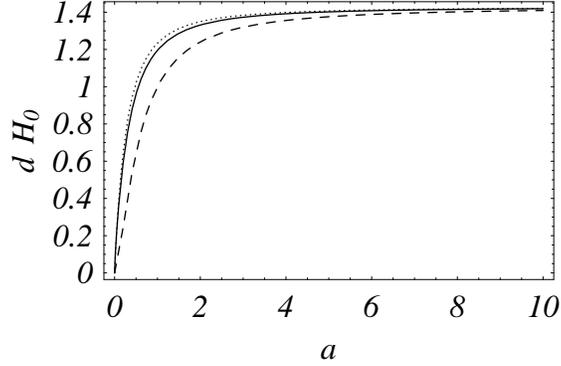}}
\caption{The evolution of various cosmological scales 
$d$ in units of $H_0^{-1}$ as functions of $a$, 
all obtained from the model described by Eqs. (9), (1) and (5)
for $r_0=3/7$.
The future event horizon is represented by the dotted line,
the scale $\mu^{-1}\sqrt{3/8\pi\kappa}$ given by (14) 
by the solid line and the Hubble distance
by the dashed line.} 
\label{fig1}
\end{figure}

One can also find that the gravitational coupling $G_N $ scales in the above
model as
\begin{equation}\label{G}
G_N=G_{N0}\frac{r_0+a^{3/2}}{r_0+1}\;,
\end{equation}
so that the minimal value for $a=0$ is $G_N = 0.3 G_{N0}$ for $r_0
= 3/7$. We also find that the ratio $\dot{G}_{N}/G_N $ is static, 
and is given by the value
\begin{equation}
\frac{\dot{G}_N}{G_N}=H_0 \frac{3}{2(1+r_0)} \;.
\end{equation}
Eqs. (16) and (17) are at least marginally consistent  with the bound on 
variation 
of the 
gravitational
coupling from primordial nucleosynthesis \cite{24}, as well as 
with the upper bound from Solar Systems experiments for the
quantity (17) \cite{25}.

Let us now study the effective dark-energy EOS. Since the matter expansion
rate is modified by a scaling of $G_N $, we need to adopt the
framework of the effective EOS (for $\rho_{\Lambda}$) 
put forward by Linder and Jenkins \cite{26}, in order to
compare our model with observations. By extracting the standard matter
contribution in the Friedmann equation, the rest is by definition the 
effective dark-energy density $\rho_{\Lambda}^{eff}$,
\begin{equation}\label{fried3}
H^2=\frac{8\pi G_{N0}}{3}(\rho_{m0}a^{-3}+\rho_{\Lambda}^{eff})\;.
\end{equation}
The effective EOS can then be defined as
\begin{equation}\label{w}
w_{eff}(a) =-1-\frac{1}{3}\frac{a}{\rho_{\Lambda}^{eff}}
\frac{d\rho_{\Lambda}^{eff}}{da} \;.
\end{equation}
One finds that the effective EOS for the present model is
\begin{equation}
w_{eff}(a) =-1-\frac{ r_0(1-a^{3/2}) }{ (r_0+a^{3/2})^2-r_0(1+r_0) } \;.
\end{equation}
In particular, the present value corresponding to $a=1$ is $w_{eff} = -1$.
In fact, the qualitative behavior of $w_{eff}$ is similar to that obtained
for the running CC models which obey the equation of continuity (4)
\cite{19}. It is easy to convince one that $w_{eff} < -1$ for small $z \; (z=a^{-1}
-1)$, but note
that arguments leading to the Big Rip no longer apply to running CC models
\cite{19, 27,27b}. The reason for having  
$w_{eff} < -1$ here lies in the modified expansion
rate for matter caused by the variable $G_N $, 
and not in the exotic nature of
$\rho_{\Lambda}$.

Let us now briefly comment on a qualitative difference between the results
presented here and those obtained in another class of generalized
holographic dark energy models introduced recently by Elizalde {\it et al.}
\cite{27b}. While we promote the Newton constant to a scale-dependent
quantity, they identify the IR cutoff with some combination of natural
IR cutoffs: the Hubble distance, the particle horizon distance, the future
event horizon, or even the length scale associated with the cosmological
constant or with the span of life of the universe 
(when the lifetime of the universe
is finite). In the models mostly considered 
within the above class of holographic dark energy,
a Big Rip sort of singularity is encountered within the effective
phantom phase. It was shown though 
\cite{27b, ht0405034, ht0405078, ht0501025} 
that quantum and quantum gravity effects might prevent evolution 
towards the Big Rip. 
In contrast, there is obviously no occurrence of the Big Rip
singularity in our solution (15), and by closer inspection of the effective
EOS (20) one sees that the effective phantom phase is only transient,
approaching $-1$ from above in the limit $t \rightarrow \infty $. An
intriguing possibility to unify the early-time inflation and late-time
acceleration of the universe within the same model of the type \cite{27b}
has emerged recently \cite{ht0506212}. Now the bottom line is to have
transitions between phantom and nonphantom phases with at least two phantom
phases corresponding to inflation at early times and acceleration at late
times. The nonphantom phase can then be identified with, e.g.,
matter-dominated era. In our model, however, a phantom-nonphantom transition  
at early time does not occur, and hence, 
without introducing additional degrees
of freedom, our model is not capable to unify early-time inflation 
with late-time acceleration of the universe.

Finally, let us consider the transition redshift $z_{*}$, which denotes a
switch of the universe from deceleration to acceleration, or equivalently,
the redshift at which $\ddot{a} = 0$ in (15). One finds $z_{*} = (r_0
/2)^{-2/3} -1$. Recent observations indicate $z_{*} < 0.72 $ at 2$\sigma $
\cite{7}. 
The presence of baryons and possibly massive neutrinos in
the matter sector of our model
enables one to make $r_0 $ as large as 0.35/0.65, but $z_{*}$ 
less than about 1.4 still cannot be obtained. 
This is definitely the weakest point of our
model. The problem here lies in the fact that the only free parameter 
of the model is $r_0
$ (which reflects the unknown $c_0 $), 
which is by observational data allowed to vary only by a very small
amount. Thus in a given setup we have no available free parameters in the
model to improve the above fit. Note also that all physical quantities do
not depend on the parameter $\kappa $.    

To this end, we make some suggestions how to modify the minimal setup 
explored
above. Obviously, the minimal setup consistent with the generalized
holographic energy density is not completely satisfactory when confronting
the most recent observational data. One way is to add matter in the
equation of continuity. Generally, we obtain
\begin{equation}
\dot{G}_{N}(\rho_{\Lambda } + \rho_m ) + G_N \dot{\rho }_{\Lambda } +
G_N (\dot{\rho }_{m} + 3H\rho_m )  = 0  \;.
\end{equation}
Note that even for the simplest case $\rho_{\Lambda} = const.$, the scale
$\mu $ is not fixed in (21). So, the scale and an additional parameter
describing deviation of the matter density from its canonical form would be
introduced as new parameters in the model. It would be interesting to see if
compatibility between  holography and observations can be achieved for some
`standard' choice for $\mu $: the inverse Hubble distance, the inverse particle
horizon or the inverse future horizon. Another suggestion would be to 
introduce
a light scalar field with its potential in the model. These would change the
equation of continuity, which now also requires the EOS of the scalar field.
We note that having two different components of the vacuum energy in the
dark-energy setup might be of some interest, see \cite{10} and 
\cite{27b}. Still, in a
running CC scenario accommodating holography such a construction is not 
appealing; a light degree of
freedom is redundant there from the beginning.

To summarize, we have considered a phenomenological scale-dependence for the 
cosmological constant as predicted by gravitational holography. The role of
the scaling parameter is played by the IR cutoff in such a manner that the
information from quantum gravity be consistently encoded in ordinary quantum
field theory. In the scenario in which also Newton's constant scales with the
same parameter, we have shown that the choice for the IR cutoff in the form
of the Hubble parameter is not phenomenologically viable, either for 
spatially flat or for curved universes. The same scenario has been shown to
have a potential to univocally fix the scaling parameter when various
variable CC models are constrained to be concordant with holography. We have
shown that in this scenario the phenomenological viability of these models is
substantially increased, although the minimal scenario is not quite
satisfactory. We have indicated that models which might have a potential to
improve the minimal scenario have much less predictive power.       

{\bf Acknowledgments. } The authors are grateful to H. \v{S}tefan\v{c}i\'c
for valuable remarks on the manuscript.
This work was supported by the Ministry of Science,
Education and Sport
of the Republic of Croatia under contract No. 0098002 and 0098011.


\begin{thebibliography}{160}
\bibitem{1} A. Cohen, D. Kaplan, and A. Nelson, Phys. Rev. Lett. 82, 4971
(1999).
\bibitem{2} J. D. Bekenstein, Phys. Rev. D7, 2333 (1973); Phys. Rev. D23,
287 (1981).
\bibitem{3} S. Weinberg, Rev. Mod. Phys. 61, 1 (1989).
\bibitem{4} G. 't Hooft, gr-qc/9310026.
\bibitem{5} L. Susskind, J. Math. Phys. (N. Y.) 36, 6377 (1995).
\bibitem{6} S. D. Hsu, Phys. Lett. B594, 13 (2004) [hep-th/0403052].
\bibitem{7} A. G. Reiss et. al., Astrophys. J. 607, 665 (2004).
\bibitem{8} P. J. Steinhardt, in ``Critical Problems in Physics'', edited by
V. L. Fitch and  Dr. R. Marlow (Princeton University Press, Princeton, N.
J., (1997).
\bibitem{9} R. Horvat, Phys. Rev. D70, 087301 (2004) [astro-ph/0404204].
\bibitem{10} B. Guberina, R. Horvat, and H. \v{S}tefan\v{c}i\'c, 
JCAP 0505, 001 (2005)
[astro-ph/0503495].
\bibitem{11} D. Pav\'{o}n and W. Zimdahl, gr-qc/0505020.
\bibitem{12} A. Babi\'c, B. Guberina, R. Horvat, and H. \v{S}tefan\v{c}i\'c,
Phys. Rev. D71, 124041 (2005) [astro-ph/0407572].
\bibitem{13} M. Li, Phys. Lett. B603, 1 (2004) [hep-th/0403127].
\bibitem{14} B. Wang, Y. Gong and E. Abdalla, hep-th/0506069.
\bibitem{15} Q-C. Huang and Y. Gong, JCAP 0408, 006 (2004)
[astro-ph/0403590]; Y. Gong, Phys. Rev. D70, 064029 (2004) [hep-th/0404030];
Q-C. Huang and M. Li, JCAP 0408, 013  (2004) [astro-ph/0404229]; K. Enqvist
and M. S. Sloth, Phys. Rev. Lett. 93, 221302 (2004); S. Hsu and A. Zee,
hep-th/0406142; K. Ke and M. Li, hep-th/0407056; A. J. M. Medved,
hep-th/0501100; B. Wang, E. Abdalla and R-K. Su, Phys. Lett. B611, 21 (2005) 
[hep-th/0404057]; F. Bauer, gr-qc/0501078; Y. S. Myung, hep-th/0412224;
hep-th/0501023; S. Nojiri and S. D. Odintsov, hep-th/0506212.

\bibitem{fot1}
A similar 
framework for holographic dark energy which allows for energy production was 
presented in \cite{16}.

\bibitem{16} Y. S. Myung, hep-th/0502128.

\bibitem{fot2}
A convenient modeling (not considered in
this paper) for the
time-dependent gravitational coupling is through a time-varying scalar
as in the framework of the Brans-Dicke theory, see \cite{gr-qc/0507010},
\cite{gr-qc/0507115} and the second reference in \cite{15}.

\bibitem{gr-qc/0507010}
H. Kim, H. W. Lee and Y. S. Myung, gr-qc/0507010.
\bibitem{gr-qc/0507115}
S. Das and N. Banerjee, gr-qc/0507115.
\bibitem{17} J. Julve and M. Tonin, Nuovo Cimento B46, 137 (1978); E. S.
Fradkin and A. A. Tseytlin, Nucl. Phys. B201, 469 (1982); A. O. Barvinsky,
Phys. Lett. B159, 269 (1985).

\bibitem{fot3}
The solution of Eq. (6) can be obtained
analytically
for $\gamma = 2$, $a = \sqrt{\eta } t + c t^2 $, where c is a constant.
Thus, albeit $\ddot{a} = const$, we find that $r$ is not a constant, being a
decreasing function of cosmic time for $\Omega_{k 0} < 0$. We thus have a
curious universe, in which a transition between matter domination and CC
domination does occur, but  in which a transition between deceleartion
and acceleration is impossible.

\bibitem{18} J. M. Overduin, F. I. Cooperstock, Phys. Rev. D58, 043506 (1998) 
[astro-ph/9805260].
\bibitem{19} P. Wang, X-He. Meng, Class. Quant. Grav. 22, 283 (2005) 
[astro-ph/0408495].  
\bibitem{20} I. L. Shapiro, J. Sola, C. Espana-Bonet, and P. Ruiz-Lapuente,
Phys. Lett. B574, 149 (2003)
[astro-ph/0303306].
\bibitem{21} C. Espana-Bonet, P. Ruiz-Lapuente, I. L. Shapiro and J. Sola,
JCAP 0402, 006 (2004)  [hep-ph/0311171].

\bibitem{fot4}
Note that $c_2$ 
is an input from the theory. On the other hand, there is no prediction for
$c_0$
whatsoever in any theory (including gravitational holography). The law
leading to (9) is a derivative one (obtained from the $\beta$-function),
thus having a natural appearance of a nonzero $c_0$. The importance of
having nonzero $c_0$ is stressed in the text below.

\bibitem{22} I. L. Shapiro and J. Sola, Phys. Lett. B475, 236 (2000) 
[hep-ph/9910462].
\bibitem{23} A. Babi\'c, B. Guberina, R. Horvat, and H. \v{S}tefan\v{c}i\'c, 
Phys. Rev.
D65, 085002 (2002) [hep-ph/0111207]; B. Guberina, R. Horvat, and H.
\v{S}tefan\v{c}i\'c, Phys. Rev. D67, 083001 (2003) [hep-ph/0211184].
\bibitem{24} See e.g. F. S. Accetta, L. M. Krauss and P. Romanelli, Phys.
Lett. B248, 146 (1990).

\bibitem{fot5}
Note that observational data
fix the present-day value for the
scale $\mu $ in (9) at $\mu_0 \sim H_0 $. However, this does not fix the
scale at
$\mu \sim H $ as long as $\rho_{\Lambda}$ from (9) is considered together
with the equation of continuity (4).

\bibitem{25} G. T. Gillies, Rep. Prog. Phys. 60, 151 (1997).
\bibitem{26}  E. V. Linder and A. Jenkins, MNRAS 346, 573 (2003)
[astro-ph/0305286]. 
\bibitem{27} R. R. Caldwell, M. Kamionkowski and N. N. Weinberg, Phys. Rev.
Lett. 91, 071301 (2003) [astro-ph/0302506].
\bibitem{27b}
E. Elizalde, S. Nojiri, S. D.
Odintsov and P. Wang, Phys. Rev. D 71, 103504 (2005) [hep-th/0502082].
\bibitem{ht0405034}
E. Elizalde, S. Nojiri, and S. D.
Odintsov, Phys. Rev. D 70, 043539 (2004) [hep-th/0405034].
\bibitem{ht0405078}
S. Nojiri and S. D. Odintsov, 
Phys. Lett. B 595, 1 (2004) [hep-th/0405078].
\bibitem{ht0501025}                                                  
S. Nojiri, S. D. Odintsov, and S. Tsujikawa, 
Phys. Rev. D 71, 063004 (2005) [hep-th/0501025].
\bibitem{ht0506212}                                                  
S. Nojiri and S. D. Odintsov [hep-th/0506212].


\end{thebibliography}
\end{document}